\documentclass{elsart} 
\usepackage{graphicx,amsmath,subfigure} 
\usepackage{amssymb} 
 
\renewcommand{\prob}{{\rm Prob}} 
\newcommand{\sign}{{\rm sign}} 
\newcommand{\avg}[1]{\left\langle{#1}\right\rangle}

\renewcommand{\l}{\left} 
\renewcommand{\r}{\right} 
  
\begin{document}

\begin{frontmatter}
\title{Coordination, intermittency and trends in generalized Minority
Games} 
\author[SMC]{A. Tedeschi\thanksref{Corr}},
\author[SMC]{A. De Martino} and
\author[SMC,ISC]{I. Giardina}
\address[SMC]{INFM SMC and Dipartimento di Fisica,
Universit\`a di Roma ``La Sapienza''\\p.le A. Moro 2, 00185 Roma
(Italy)}
\address[ISC]{CNR ISC, via dei Taurini 19, 00185 Roma (Italy)}
\thanks[Corr]{Corresponding author. Email: tedeschi@roma1.infn.it}

\begin{abstract}
The Minority Game framework was recently generalized to account for
the possibility that agents adapt not only through strategy selection
but also by diversifying their response according to the kind of
dynamical regime, or the risk, they perceive. Here we study the
effects of this mechanism in different information structures. We show
that both the stationary macroscopic properties and the dynamical
features depend strongly on whether the information supplied to the
system is exogenous (`random') or endogenous (`real'). In particular,
in the latter case one observes that a small amount of herding
tendency suffices to alter the collective behavior dramatically. In
such cases, the dynamics is characterized by the creation and
destruction of trends, accompanied by intermittent features like
volatility clustering.
\end{abstract}

\begin{keyword}
Minority games, risk, trend-followers, contrarians, volatility
clustering
\end{keyword}

\end{frontmatter}

\section{Introduction} 
 
The Minority Game (MG) and related models allowed to elucidate many
aspects of the critical behavior of systems of heterogeneous inductive
agents, with special emphasis on financial markets, by addressing
directly the interplay between microscopic behavior and macroscopic
properties -- including fluctuations, market impact, predictability
and efficiency -- in rather elementary and often exactly solvable
settings \cite{book,jjh,cool}. Besides providing a surprisingly rich
description of the phase structure of an idealized speculative market
\cite{mod}, MG-based models are also able to reproduce to some extent
the empirical price statistics (the so-called `stylized facts'
\cite{ms}) of financial markets \cite{gcmg,farmer,hui}. However, the
emergence of the peculiar dynamical regimes characterizing real
systems, with intermittent fluctuation outbursts (volatility
clustering) and the formation and destruction of trends, is only
partially captured by standard MGs. This is in part due to the fact
that the critical window where empirical phenomenology is observed
typically shrinks as the system size increases, indicating that
stylized facts would require a significant amount of fine tuning of
the parameters and ultimately marginal efficiency
\cite{gcmg}. Moreover, the dynamics of these models usually displays a
strong dependence on both the disorder sample and the initial
conditions \cite{chal}. It would therefore be desirable to devise a
simple microscopic mechanism that is able to generate a realistic
dynamical structure in a robust way while preserving the fundamental
physical content of MGs.

Several studies in financial market microstructure suggest that these
peculiarities emerge from the coexistence of different types of
traders, and point the attention especially on `fundamentalists' and
`chartists' \cite{frfr,kir,set,lm,gb,Chiarella,SantaFe,Flo}. The
former trade on the spread between the actual and the expected price
(the `fundamental') trying to profit from fluctuations; the latter
look for trends in the price history and follow the expected
drift. The common picture has it that in market phases dominated by
fundamentalists the price follows the fundamental, whereas when
chartist pressure prevails trends deviating from the fundamental
(bubbles) occur. Moreover, switches between the two phases, during
which agents revise their expectations over a certain time span, are
possibly accompanied by activity outbursts.

The expectations of agents in MGs are implicitly encoded in the
reinforcement term of the learning dynamics. They are usually fixed:
agents behave as fundamentalists in MGs and as chartists in Majority
Games \cite{matteo,as,f,km}. To implement the mechanism discussed
above means therefore to modify the learning process so as to allow
agents to change their expectations, and hence their character, in
time according to the market conditions they perceive. One possible
way to do this was introduced in \cite{prerc}. In a nutshell, the idea
is that when price movements are small agents should try to look for
profit opportunities in emerging trends while when large price changes
occur they should perceive a greater risk as the market behaves more
chaotically, and return to a more cautious conduct. The passage from
one attitude to the other is provided by a parameter that tunes the
agents' risk-sensitivity -- or the trade-off between expected profit
and perceived risk -- so that when price movements exceed a given
threshold the risk perceived by agents is large and they play a MG
(that is, they select strategies that effectively try to follow the
fundamental) preferentially. When coupled to a random external
information structure, this mechanism was shown to lead to the
formation of `heavy tails' in the distribution of returns in a large
`critical' window where a crossover from a fundamentalist-dominated to
a trend-followers dominated market takes place.

Here we discuss this setting further by considering a somewhat simpler
but more transparent model in order to address the role of the
information structure, particularly on the system dynamics. This is an
especially crucial point in analyzing the interaction between
trend-followers and contrarians, since trend-following behavior is
expected to introduce a strong bias in the endogenous information
dynamics. We will indeed see that a small trend-following attitude is
sufficient to drastically change both the typical stationary-state
macroscopic properties and the single-sample dynamical properties with
respect to the random information case. We will proceed by adding
increasingly complex information structures to the simplest possible
model (without information), which is introduced in Section 2. In
Section 3 the case of random external information is considered while
in Section 4 we address the case of `real' endogenous
information. Finally, we formulate some concluding remarks in Section
5.

\section{The simplest model (without information)}

Let us consider the following setup. Each of $N$ agents
($i=1,\ldots,N$) must decide whether to buy ($a_i(t)=1$) or sell
($a_i(t)=-1$) at each time step $t=0,1,\ldots$. The success of agent
$i$ at time $t$ is measured by the payoff function
\begin{equation}
\pi_i(t)=a_i(t)A(t)f[A(t)]
\end{equation}
where $A(t)=\sum_i a_i(t)$ is the aggregate bid (or `excess demand',
which serves as a proxy for price movements). The function $f$ encodes
the type of game being played and, implicitly, the agents'
expectations. If $f$ is a constant, agents are either playing a
Majority Game (for $f>0$, $\pi_i>0$ if $i$ acts according to the
majority) or a Minority Game (for $f<0$, $\pi_i>0$ if $i$ acts
according to the minority). Agents thus behave as trend-followers in
the former case and as fundamentalists in the latter. We want to
address a more general case in which agents are able to modify their
character depending on the size of market fluctuations. We assume that
they behave as trend-followers (resp. fundamentalists) when price
movements are small (resp. large) and the perceived risk is small
(resp. large). The simplest function for our scope is perhaps
\begin{equation}
f(x)=\chi\l(|x|<L\r)-\chi\l(|x|>L\r)\label{nostra}
\end{equation}
where $\chi(B)=1$ if $B$ is true (and $0$ otherwise) and $L$ is a
threshold, so that then $|A(t)|<L$ agents perceive the game as a
Majority Game, whereas for $|A(t)|>L$ they revert to a Minority
Game. For the sake of simplicity, in what follows we assume that
$L=O(N)$.

We imagine that in order to make their decisions agents employ the
following iterated probabilistic rule ($a\in\{-1,1\}$):
\begin{gather}
\prob\{a_i(t)=a\}=\frac{e^{a U_i(t)}}{2\cosh U_i(t)}\\
U_i(t+1)-U_i(t)=\Gamma A(t) f[A(t)]/N
\end{gather} 
where $\Gamma>0$ is a constant (the `learning rate' of agents) and
$U_i(0)$ are independent, identically distributed quenched random
variables with zero mean and variance 1 for
$i=1,\ldots,N$~\footnote{In principle, the parameters $L$ and $\Gamma$
and the function $f$ itself could be different for different
agents. We however drop this source of heterogeneity in order to focus
on the effects induced by risk sensitivity. On the other hand, it is
obvious that the heterogeneity of initial conditions $U_i(0)$ cannot
be dropped, since $U_i(t)-U_i(0)$ is independent on $i$ for all $t$
and the dynamics would be completely trivial if the starting point was
the same for everybody.}, and consider the properties of the steady
state of the dynamics. We will see that, in spite of its extreme
simplicity, this model is sufficient to capture one of the key
features induced by this scheme.

\begin{figure}[t]
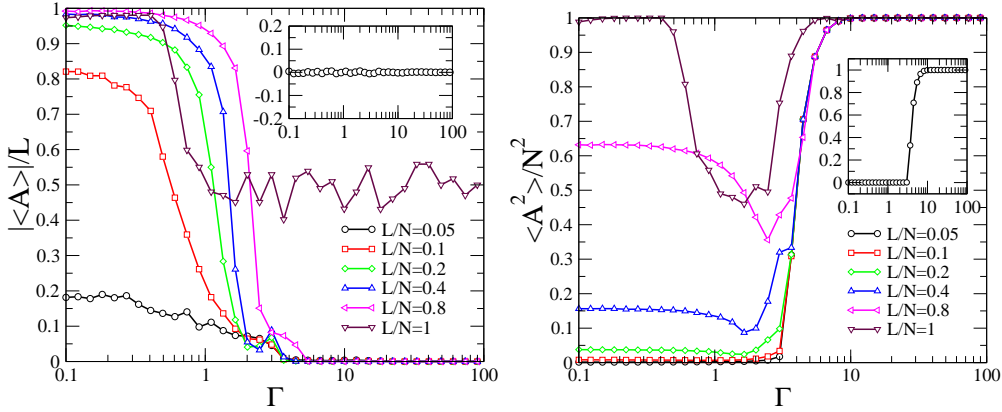

\begin{center}
\includegraphics*[width=0.47\textwidth]{avgA_simple.eps}
\includegraphics*[width=0.47\textwidth]{vola_simple.eps}
\caption{\label{one}$|\avg{A}|/L$ (left) and $\avg{A^2}/N^2$ (right)
versus $\Gamma$ for different values of $L$. Insets: $\avg{A}$ vs
$\Gamma$ (left) and $\avg{A^2}/N^2$ (right) versus $\Gamma$ for a pure
MG, corresponding to $L=0$ (system with $N=1000$, averages over $200$
disorder samples per value of $\Gamma$).}
\end{center}
\end{figure}

\begin{figure}[t]
\begin{center}
\includegraphics*[width=0.7\textwidth]{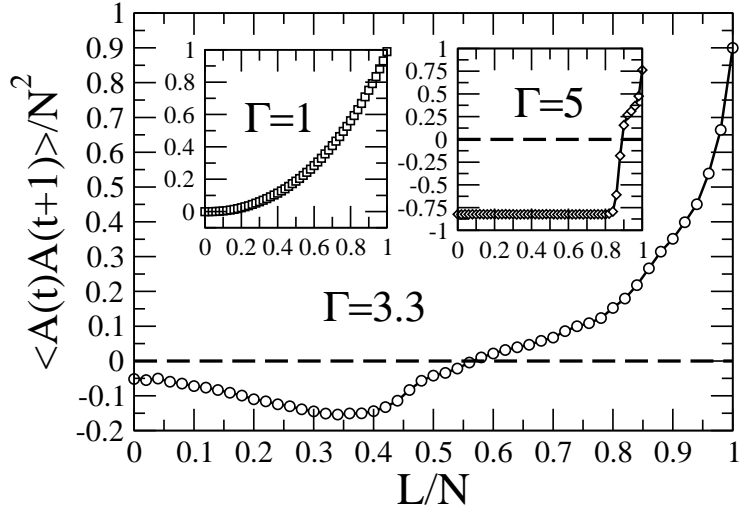}
\caption{\label{two}$\avg{A(t)A(t+1)}/N^2$ versus $L/N$ for $\Gamma=1$
(left inset), $\Gamma=3.3$ (main plot) and $\Gamma=5$ (right inset)
(system with $N=1000$, averages over $200$ disorder samples per value
of $L/N$)}
\end{center}
\end{figure}

Let us see, for a start, the behavior of the time average of the
excess demand $\avg{A}$ (where the average is taken in the steady
state of the dynamics). In a Majority Game ($f(x)=1$ or $L\geq N$) one
finds $\avg{A}\neq 0$, implying that one of the two possible actions
is systematically preferred (a trend), while $\avg{A}=0$ in a pure MG
($f(x)=-1$ or $L=0$). Intuition suggests that for a sufficiently small
$L$ the latter scenario will dominate, since agents will be extremely
risk-sensitive. On the other hand, for a sufficiently large $L$ agents
will be more risk-prone and a Majority-Game scenario is expected. The
crossover between these two regimes is displayed in Fig. \ref{one}.
One observes that for small $\Gamma$ trends are formed (as in a
Majority Game) even for small values of $L$, wheres for large $\Gamma$
the typical excess demand returns to zero (as in a MG) for all
$L<N$. It is also instructive to look at the crossover in the second
moment $\avg{A^2}$, recalling that in a Majority Game, $\avg{A^2}$ is
of order $N^2$ for any $\Gamma$, while in a pure MG (see also insets
in Fig. \ref{one}) $\avg{A^2}$ is of order $N$ for small $\Gamma$ and
becomes of order $N^2$ for large enough $\Gamma$.  Finally, a deeper
insight may be obtained from the correlation function
$\avg{A(t)A(t+1)}$ (see Fig. \ref{two}).
For intermediate $\Gamma$ excess demands are negatively correlated for
low $L$ whereas the correlation turns positive when $L$ increases,
that is as agents become less and less risk sensitive. The former
regime is characteristic of Minority Games, where price increments are
one-step anti-correlated because agents effectively act so as to
compensate price increments in one direction with subsequent
increments in the other direction. The latter regime is instead
typical of trends and Majority Games.

This coexistence of MG-like and of Majority-Game-like features is at
odds with mixed models in which each agent is either a fundamentalist
or a chartist and is not allowed to pass from one group to the other
\cite{matteo,mmm}. In that case, the macroscopic properties are
essentially determined by the larger group (be it fundamentalists or
trend-followers), whose expectations are fulfilled. Here, the risk
sensitivity determines which group of traders dominates and mixed
phenomenology occurs in a range of values of the learning rate
$\Gamma$. While the detailed physical picture becomes more articulate,
this conclusion actually extends to models endowed with more
complicated information structure.

\section{Case of random external information} 
 
Let us move over to the usual MG setting, namely a system of $N$
agents who at each time step $t$ must formulate a binary bid
(buy/sell) based on some public information pattern $\mu(t)$. One such
pattern is given to agents at every time step; it is assumed that the
number of possible patterns is $P$ and that $P$ scales linearly with
$N$. The relevant control parameter is indeed the relative number of
information patterns $\alpha=P/N$. In order to translate informations
into bids each agent disposes of $S$ strategies, each one predicting
the outcomes $\boldsymbol{a}_{ig}=\{a_{ig}^\mu\}$ ($i=1,\ldots,N$;
$g=1,\ldots,S$; $\mu=1,\ldots,P$), and aims at selecting, at each time
step, the one that delivers him the highest expected profit. If we
denote this optimal strategy as $g_i(t)$, the agent's bid at time $t$
is then given by $a_{i g_i(t)}^{\mu(t)}$. As usual, we assume that
strategies have quenched random components drawn independently from
$\{-1,1\}$ with equal probability. In this work, we focus on the case
$S=2$.

The single-agent dynamics is defined by the following rules:
\begin{gather}
g_i(t)=\text{arg }\max_g U_{ig}(t)\nonumber\\
A(t)=\frac{1}{\sqrt{N}}\sum_i a_{ig_i(t)}^{\mu(t)}\\
U_{ig}(t+1)-U_{ig}(t)=a_{ig}^{\mu(t)}A(t)f[A(t)]\nonumber
\end{gather}
where each $U_{ig}$, called the `score' in MG jargon, measures the
performance of strategy $g$ of agent $i$. At each time step, every
agent chooses his best-performing strategy as the one with the highest
score and formulates the corresponding bid. Subsequently, bids are
aggregated into the excess demand $A(t)$, that we have now normalized
for future convenience, and scores are
updated. It is assumed here that scores are initialized at time $t=0$
in such a way that $U_{i1}(0)-U_{i2}(0)=0$ for all $i$ and $g$
(`unbiased' or `flat' initial conditions). We will not address here
the important and subtle issues of if and how the steady state changes
when a non-zero initial bias is used.

In this Section, the information structure is assumed to correspond to
the random external case: $\mu(t)$ is an integer drawn randomly and
independently at each time step from $\{1,\ldots,P\}$ with uniform
probability \cite{cava}. In this case the model is Markovian and the
information dynamics covers uniformly the state space
$\{1,\ldots,P\}$. As for $f$, we can borrow the recipe employed in the
previous section and set:
\begin{equation}
f(x)=\chi\l(|x|<\eta\r)-\chi\l(|x|>\eta\r)\label{nostra2}
\end{equation}
where $\eta>0$ is a tunable constant (notice that $A(t)$ is of order
$1$ in this case). Different choices are also possible. For example,
linearizing (\ref{nostra2}) one obtains
\begin{equation}
f(x)=\eta-|x|
\end{equation}
which is equally well suited for our purposes. The main difference
with the piecewise constant case lies in the fact that payoffs (which,
we remind, are proportional to $A(t)f[A(t)]$) are a non-linear
function of the aggregate bid. Such non-linear choices, one of which
was considered in \cite{prerc}, lead to the emergence of a clear
non-Gaussian statistics for $A(t)$ at small values of $\alpha$. Here
we wish to concentrate on the role of information structures and
therefore we may restrict ourselves to the somewhat simpler form
(\ref{nostra2}).

We will analyze the macroscopic properties in the steady state using
$\alpha$ and $\eta$ as control parameters. Our attention will be
mostly focused on: (a) the volatility $\sigma^2=\avg{A^2}$ measuring
the magnitude of global fluctuations (notice that $\avg{A}=0$ by
construction); (b) the `predictability'
\begin{equation}\label{pred_r}
H=\frac{1}{P}\sum_\mu\avg{A|\mu}^2
\end{equation}
where $\avg{A|\mu}$ stands for the time average of $A$ conditioned on
the occurrence of the pattern $\mu$, quantifying the presence of
exploitable information (if $H\neq 0$ the minority action can be
statistically predicted on the basis of the information pattern alone
at least for some $\mu$); and finally (c) the (normalized) one-step
autocorrelation function $D=\avg{A(t)A(t+1)}/\sigma^2$, which
indicates the dominating component of the market (if $D>0$ returns are
positively correlated and trend-followers dominate). Other observables
of interest will be defined in due course.
  
\begin{figure}[t]
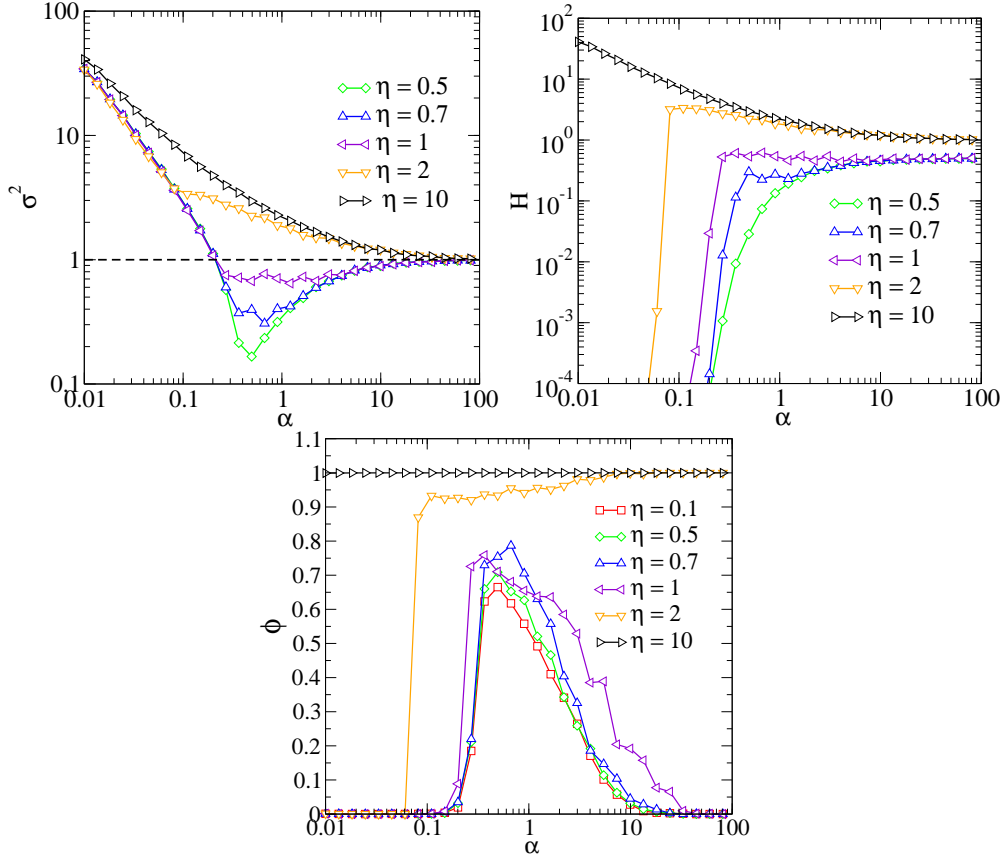

\begin{center} 
\includegraphics*[width=0.47\textwidth]{Theta_s2.eps} 
\includegraphics*[width=0.47\textwidth]{Theta_H.eps} 
\includegraphics*[width=0.47\textwidth]{Theta_phi.eps} 
\caption{\label{M1_s2} Stationary volatility $\sigma^2$,
predictability $H$ and fraction of frozen agents $\phi$ as a function
of $\alpha$ for different values of $\eta$. Simulations performed with
$\alpha N^2=16000$, with averages over at least 200 disorder samples
per point.Every sample corresponds to a particular realization of the
strategies. Numerical results for values of $\eta$ lower or larger
than those displayed are not visibly different.}
\end{center} 
\end{figure}

\begin{figure}[t]
\begin{center}
\includegraphics*[width=0.7\textwidth]{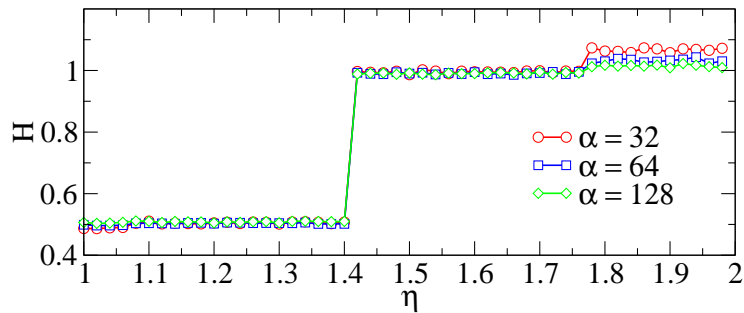} 
\caption{\label{H_vs_b}Stationary predictability as a function of
$\eta$ for different large values of $\alpha$. Simulations performed
with $\alpha N^2=16000$, with averages over 100 disorder samples per
point.}
\end{center}
\end{figure}

\begin{figure}[t]
\begin{center}
\includegraphics*[width=0.7\textwidth]{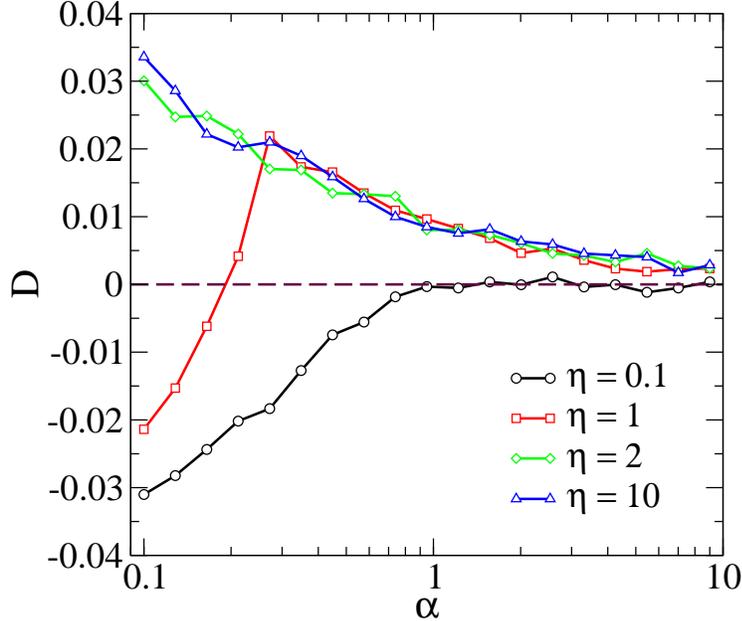}
\caption{\label{Theta_D}Normalized one-step autocorrelation function
$D$ as a function of $\alpha$ for different values of $\eta$.
Simulations performed with $\alpha N^2=16000$ and averages over 100
disorder samples per point.}
\end{center}
\end{figure}

Numerical results are shown in Fig. \ref{M1_s2}. 
One sees that the magnitude of fluctuations increases smoothly with
$\eta$, as one passes from a MG-like regime where $\sigma^2$ can be
smaller than the random-trading value 1, signaling a high degree of
coordination among agents, to a Majority-Game-like regime (large
$\eta$) where $\sigma^2>1$. Correspondingly, the system displays a
transition between an unpredictable (or `symmetric', low $\alpha$)
regime with $H=0$ to a predictable (or `asymmetric', high $\alpha$)
when $\eta$ is sufficiently small, similarly to what happens in the
MG. Notice that the critical point $\alpha_c$ below which $H=0$ shifts
(continuously) to smaller values as $\eta$ increases. Clearly, as the
agents' risk threshold grows their tendency to behave like
trend-followers increases and, by herding, they produce more and more
exploitable information ($H$ increases with $\eta$) even for smaller
systems. Note that upon increasing $\eta$ further, one still observes
a (sharp!) MG-like transition at low $\alpha$, but for high $\alpha$
the predictability tends to $1$, as in the Majority Game. The change
from one regime to the other at large $\alpha$ is apparently a
threshold phenomenon in $\eta$ (see Fig. \ref{H_vs_b})\footnote{A
similar, though less sharp threshold phenomenon was found for the
different though conceptually similar model of \cite{prerc}.}.
The coexistence of the competing tendencies can also be seen from the
fraction of `frozen' agents (namely, agents for which
$|U_{i1}(t)-U_{i2}(t)|\to\infty$ as $t\to\infty$, so that they use
only one of their strategies in the stationary state), $\phi$. For
small $\eta$, one finds a pure Minority Game; for intermediate $\eta$,
$\phi$ jumps from the small-$\alpha$ value of $0$ (as in the MG) to
the large-$\alpha$ value of $1$ (as in the Majority Game, where all
agents ultimately freeze \cite{km,mmm}). Notice that as $\eta$
decreases $\phi$ also decreases, which implies that when agents become
more risk-sensitive it becomes more difficult for them to identify an
optimal strategy. Coming to the the autocorrelation function $D$ (see
Fig. \ref{Theta_D}), it shows the coexistence of Minority-like and
Majority-like features for intermediate values of $\eta$ once more.
For small $\eta$, $D$ is negative and the dynamics is completely
dominated by anticorrelations (i.e. by contrarians). As $\eta$
increases positive correlations appear for large $\alpha$, and for
large $\eta$ the system is dominated by trend-followers.

In summary, one can say that adding a certain risk-tendency at the
microscopic level leads to a loss of global efficiency and that for
intermediate values of $\eta$ MG-like features coexist with
Majority-Game-like features, the former prevailing at low
$\alpha$. This is again in sharp contrast with the scenario emerging
from mixed Majority-Minority Games, where the expectations of the
larger group (be it fundamentalists or trend-followers) are fulfilled
at all $\alpha$ and where the macroscopic properties are essentially a
linear combination of those of the pure models \cite{mmm}.

It is important to notice that even when the competition between
risk-aversion and profit-maximization is not too strong, for instance
in the MG-like phase, the dynamics acquires several non trivial
traits. 
\begin{figure}[t]
\begin{center}
\includegraphics*[width=0.7\textwidth]{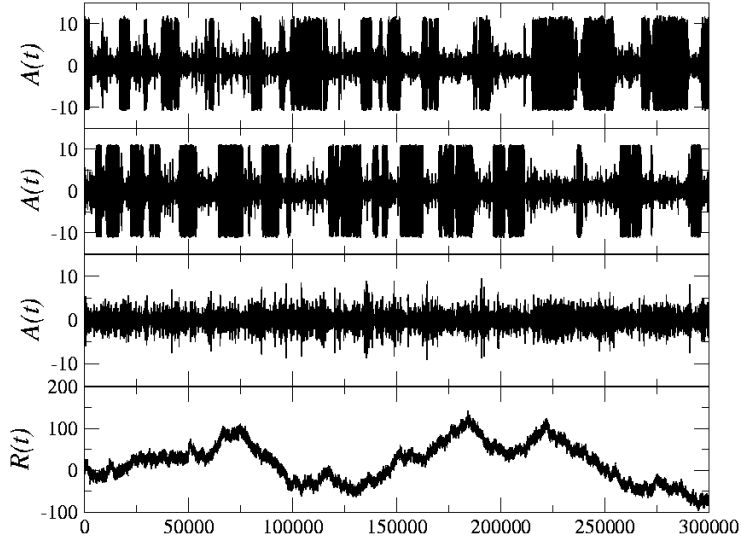} 
\caption{\label{ss} Excess demand $A(t)$ as a function of the time
elapsed from equilibration conditioned on the presence of three
different information patterns (top down) for a single sample with
$\eta=0.5$ and $\alpha=0.05$. Bottom figure: price time series for the
same realization. System of $N=565$ agents.}
\end{center}
\end{figure} 
In Fig. \ref{ss} we report the results from a single realization of
the time series of the excess demand $A(t)$ in correspondence with
three different information patterns, and the time series of the
`price' $R(t)=\sum_{\ell\leq t} A(\ell)$, both for small $\alpha$,
that is, where MG-like features are predominant. One can see that
excess demand fluctuations at fixed $\mu$ behave intermittently:
periods of high volatility are followed by periods of low
volatility. At the same time, the profile of $R(t)$ shows the
formation of well-defined trends.
 
\section{Case of endogenous information} 

We now move to the case in which the information pattern $\mu(t)$
encodes in its binary representation the string of the last $m$ losing
actions (the `history') of the market $\sign[A(t-\ell)]$
($\ell=1,\ldots,m$) so that $P=2^m$. The information dynamics in this
case is deterministic and reads
\begin{equation}
\mu(t+1)=
\begin{cases} 
\l[2\mu(t)+1\r]\text{mod $P$}&\text{if $A(t)>0$}\\ 
\l[2\mu(t)\r]\text{mod $P$}&\text{if $A(t)<0$} 
\end{cases}
\end{equation}
Trend-following behavior is expected to influence the macroscopic
properties rather strongly, because of the bias trends would impose on
the resulting history dynamics. We therefore have to take into account
the frequency $\rho(\mu)$ with which each string $\mu$ is generated in
the steady state and modify the definition of the predictability
(\ref{pred_r}) as
\begin{equation}
H=\sum_\mu\rho(\mu)\avg{A|\mu}^2
\end{equation}
As a measure of the bias (or of the information content) of the
frequency distribution one typically employs the entropy
\begin{equation}
S=-\sum_\mu\rho(\mu)\log_P\rho(\mu)
\end{equation}
which is normalized in such a way that for the random case discussed
above, $\rho(\mu)=1/P$ and $S=1$.
 
\begin{figure}[t]
\begin{center} 
\includegraphics*[width=0.7\textwidth]{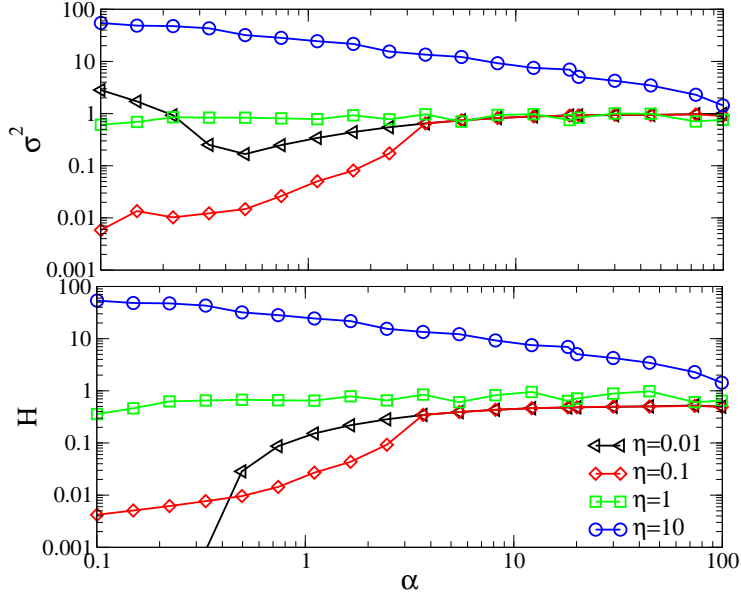} 
\caption{\label{s2_H}Volatility (top) and predictability (bottom) as a
function of $\alpha$ for different values of $\eta$. Simulations
performed with $\alpha N^2=30000$, with averages over 200 disorder
samples per point.}
\end{center}
\end{figure}
\begin{figure}[t]
\begin{center} 
\includegraphics*[width=0.7\textwidth]{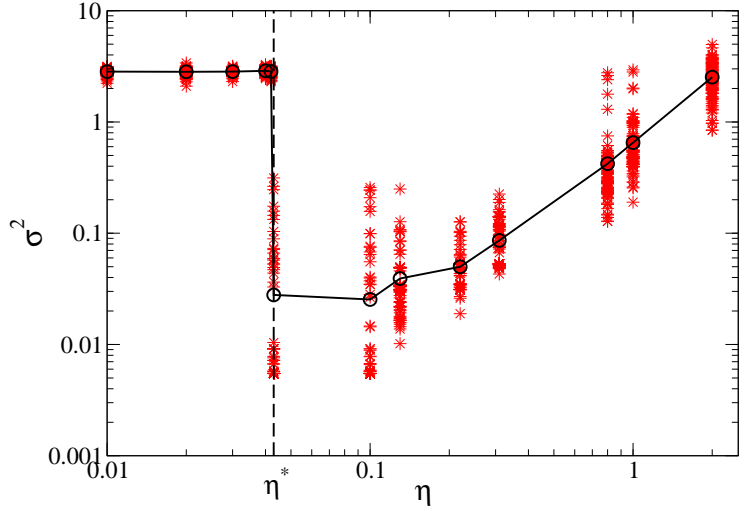} 
\caption{\label{s2b}Volatility as a function of $\eta$ for
$\alpha=0.1$. Circles correspond to averages over 200 disorder
samples, stars correspond to the values obtained in different
samples. System with $N=547$. The dashed vertical line marks the value
$\eta^*$ corresponding to the minimum possible value of $|A(t)|$.}
\end{center}
\end{figure}

 
Numerical results for the volatility $\sigma^2$ and the predictability
$H$ (see Fig. \ref{s2_H}) show that already for small values of $\eta$
(e.g. $\eta=0.1$) the behavior deviates greatly from the one found in
the case of exogenous information. In fact, while there is no evidence
of a symmetric regime, fluctuations for small $\alpha$ are much
smaller than in the MG-regime. A detailed analysis of the volatility
as a function of $\eta$ for small $\alpha$ (see Fig. \ref{s2b})
suggests that as soon as agents allow for a small amount of
risk-proneness fluctuations decrease sharply, though their value may
vary significantly from sample to sample. It is important to notice
that since $N$ is odd, the minimum possible value of $|A(t)|$ is
$\eta^*=1/\sqrt{N}$ (which corresponds to $\eta^*\simeq 0.043$ for the
simulations of Fig. \ref{s2b}). For $\eta<\eta^*$ agents are always
playing a MG which leads to large fluctuations in the low
$\alpha$-phase, given the flat initial conditions. As soon as the risk
thereshold exceeds $\eta^*$ and agents have a chance to herd
fluctuations become significantly smaller. This remarkable effect,
together with the fact that the predictability is small, indicates
that relatively efficient states can be reached. As $\eta$ increases
further $\sigma^2$ increases smoothly until it eventually reaches its
standard Majority-Game value.
(Note that the next smallest possible values of $|A(t)|$ are
$3/\sqrt{N}$, $5/\sqrt{N}$ etc.; for $\eta$ in-between these values
the stationary volatility is roughly constant). Thus, we see that the
effects due to herding are much more pronounced in the presence of
real histories with respect to the case of random information. Loosely
speaking, one could say that a small amount of greediness at the
microscopic level may turn out to have positive effects at the
macroscopic level. The toll to pay is the reduction of informational
efficiency. This is yet another proof of the fact that the interplay
between these two properties may be considerably subtle in these
systems. However larger risk thresholds lead to a serious loss of
global efficiency. For large $\alpha$, instead, the model behaves as a
pure MG. As $\eta$ increases, the pressure of trend-followers gets
stronger and the model acquires more and more the character of a
Majority Game.

\begin{figure}[t]
\begin{center}
\includegraphics*[width=0.7\textwidth]{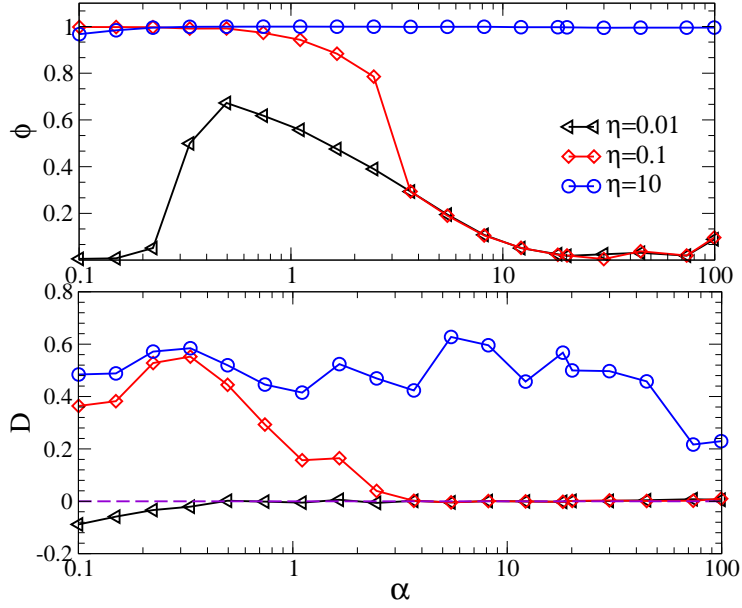} 
\caption{\label{phi_D}Fraction of frozen agents (top) and normalized
one-step autocorrelation function (bottom) as a function of $\alpha$
for different values of $\eta$. Simulations performed with $\alpha
N^2=30000$, with averages over 200 disorder samples per point.}
\end{center}
\end{figure}

Analyzing the fraction of frozen agents (see Fig. \ref{phi_D}) one
sees that for small $\alpha$ almost all agents are frozen at the
interesting values of $\eta$, so that even individual agents actually
profit from a small greediness.
Surprisingly, however, $\phi$ tends to the MG-like behavior when
$\alpha$ increases. This is in striking contrast to the observations
made for the model with exogenous information, in which MG-like
features prevail at small $\alpha$ when coexisting with Majority-type
of features. The results for the autocorrelation function $D$ confirm
that indeed a small $\eta$ is sufficient to induce strong herding
effects for small $\alpha$, at odds with the previous case.

\begin{figure}
\begin{center} 
\includegraphics*[width=0.7\textwidth]{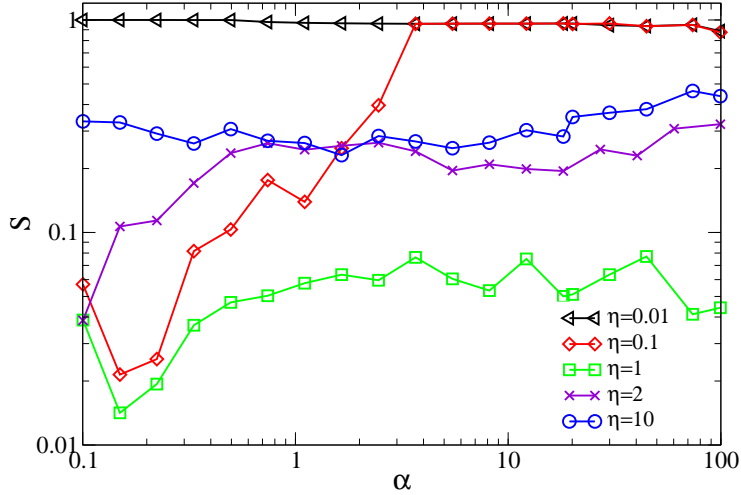} 
\caption{\label{S}Entropy as a function of $\alpha$ for different
$\eta$.  Simulations performed with $\alpha N^2=30000$, with averages
over 200 disorder samples per point.}
\end{center}
\end{figure}

\begin{figure}[t]
\begin{center} 
\includegraphics*[width=0.7\textwidth]{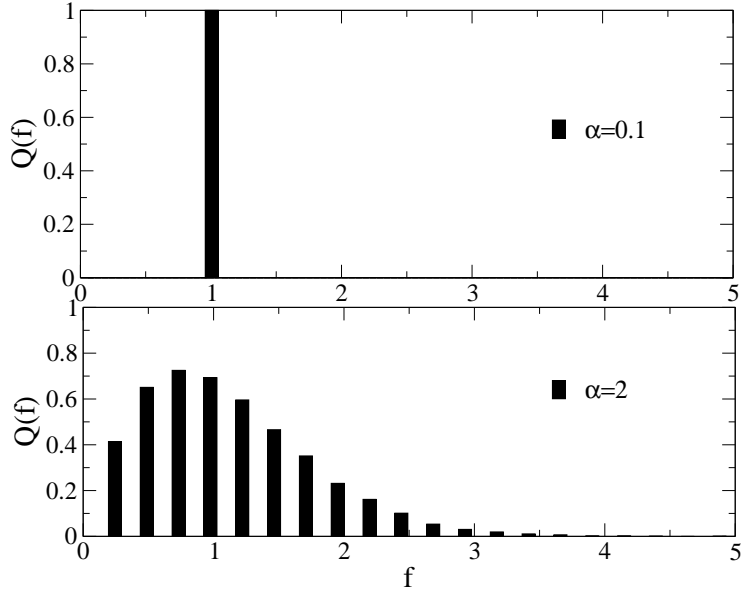}
\caption{\label{p_rho}Relative distribution of frequencies $Q(f)$ for
$\eta=0.01$ at $\alpha=0.1$ (top) and $\alpha=2$ (bottom). Simulations
performed with $\alpha N^2=30000$, with averages over 100 disorder
samples per point.}
\end{center}
\end{figure}

\begin{figure}[t]
\begin{center} 
\includegraphics*[width=0.9\textwidth]{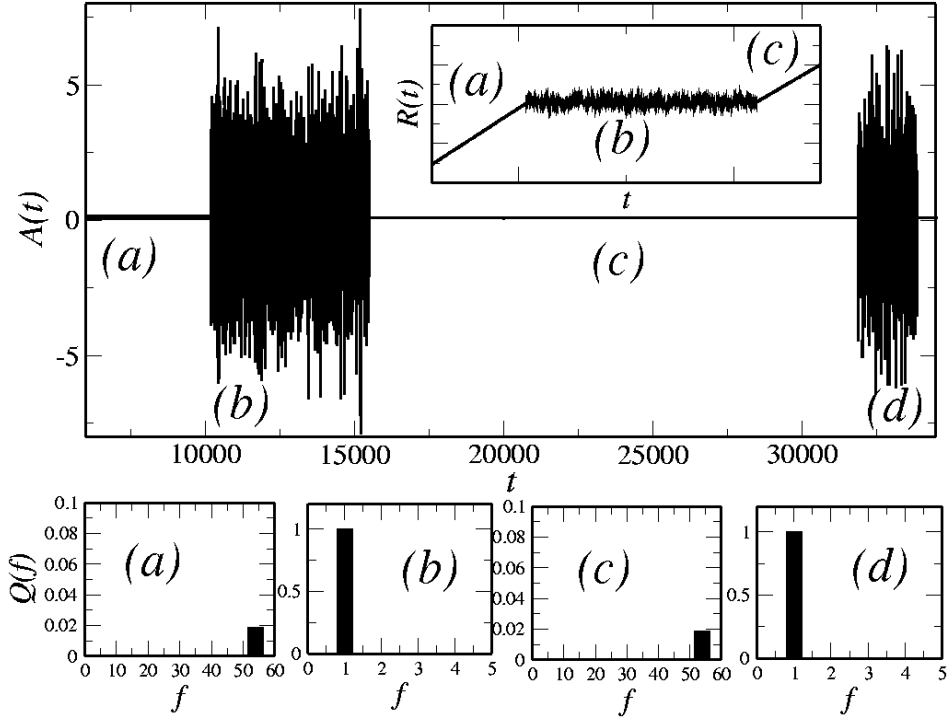} 
\caption{\label{A_rho}Single sample excess demand as a function of the
time elapsed from equilibration for $\eta=0.06$ (top) and
$\alpha=0.1$. Relative frequency distribution $Q(f)$ for the same
realization in the different regimes (bottom). Inset: `price' $R(t)$
as a function of time.  }
\end{center}
\end{figure}

Let us now look at the history dynamics. The entropy $S$ is reported
in Fig. \ref{S}.
First of all, for sufficiently small $\eta$ the scenario of a pure MG
should be reproduced, where $S=1$ for $\alpha<\alpha_c\simeq 0.34$ and
$S<1$ (slightly) for $\alpha<\alpha_c$. This is indeed the case, as
can be seen even from Fig. \ref{p_rho}, where we plot the steady-state
distribution of history frequencies relative to the uniform case:
\begin{equation}
Q(f)=\frac{1}{P}\sum_\mu\delta[f-P\rho(\mu)]
\end{equation}
(if $\rho(\mu)=1/P$ for all $\mu$, $Q(f)$ is a delta-distribution at
$f=1$). As $\eta$ increases the entropy drops seriously for small
$\alpha$ as herding trivializes the history dynamics. For large
$\alpha$, again, the MG behavior is recovered. As $\eta$ increases and
trend-following becomes more and more preferred, $S$ consistently
tends to be much smaller than $1$ as only a small fraction of
histories are generated per sample. However, a glance at the single
sample behavior suffices to understand that the dynamics is much more
complex than the entropy would and could tell (see Fig. \ref{A_rho}). 
For small $\eta$ and small $\alpha$ one distinctly observes volatility
clustering {\it in the time series of $A(t)$}. Periods of low
volatility correspond to a trend, as shown by the distribution $Q(f)$
relative to those intervals. Here, the game has the character of a
Majority Game and trend-followers dominate. Periods of high volatility
correspond instead to a chaotic dynamics where the history frequency
distribution is uniform. Here, the model behaves like a Minority Game
and fundamentalists dominate. The switch from one regime to the other
is essentially driven by random events. Indeed, the duration of the
activity outbursts, as well as that of trends, is arbitrary and may
vary strongly from sample to sample. The sharpness of the switch is
instead simply due to the particular form of the function $f$ we have
chosen, as well as to the value of $\eta$. In fact, as one increases
$\eta$, the magnitude of fluctuations in the Majority-like
periods increases and the different regimes tend to merge.
 
\section{Conclusion} 

The simple microscopic mechanism introduced in our model,
when coupled to real information, determines significant effects in the
stationary macroscopic properties and produces realistic dynamical features
such as volatility clustering. These results, together
with the observations made in \cite{prerc} on the statistics of price
changes, show that  generalized MGs reveal a remarkably rich and realistic
behavior which has been only partially uncovered so far. By all means, we
believe that the statistical mechanics of systems of interacting
trend-followers and contrarians constitute an extremely challenging
problem for physicists and are definitely worth more detailed
investigations. Of course, an analytical solution would be
welcome. To conclude we would like to indicate an  issue that is, in our
opinion, particularly interesting, namely seeing how this
scenario would change if the risk-threshold were allowed to fluctuate
in time. The  most intuitive way to do that
is perhaps to couple the threshold's dynamics
to the system performance.  A possible microscopic mechanism could be the
following. When $\eta$ is large a high volatility is to be expected as
agents are more likely to behave as trend-followers. As a consequence,
they should likely reduce their threshold since the market is risky;
however, for small $\eta$ fundamentalists are expected to dominate and
the game should acquire a Minority character. Hence the predictability
will be smaller and there will be less profit opportunities. Agents
may then decide to adopt a larger threshold to seek for convenient
speculations on a wider scale. If these two competing effects are
appropriately described by an evolution equation for $\eta$, the
system should self-organize around an `optimal' value of the risk
threshold. For all practical purposes, the model discussed in this 
work assumes that such a
time evolution takes place on time scales much longer than those over
which trading occurs.


\begin{thebibliography}{99}
\bibitem{book}D Challet, M Marsili and YC Zhang, {\it Minority Games}
(Oxford University Press, 2004)
\bibitem{jjh}NF Johnson, P Jefferies and PM Hui, {\it Financial market
complexity} (Oxford University Press, 2003)
\bibitem{cool}ACC Coolen, {\it The mathematical theory of Minority
Games} (Oxford University Press, 2004)
\bibitem{mod}D Challet, M Marsili and YC Zhang, Physica A {\bf 276}
284 (2000)
\bibitem{ms}RN Mantegna and HE Stanley, {\it An introduction to
econophysics} (Cambridge University Press, 1999)
\bibitem{gcmg}D Challet and M Marsili, Phys. Rev. E {\bf 68} 036132
(2003)
\bibitem{farmer}JD Farmer, Industrial and Corporate Change {\bf 11}
895 (2002)
\bibitem{hui}P Jefferies, ML Hart, PM Hui and NF Johnson,
Int. J. Theor. Appl. Finance {\bf 3} 3 (2000)
\bibitem{chal}D Challet, A De Martino and M Marsili, Physica A {\bf
338} 143 (2004)
\bibitem{frfr}JA Frankel and KA Froot, Am. Econ. Rev. (Pap. Proc.) {\bf 80} 181 (1990)
\bibitem{kir}A Kirman, Quart. J. Econ. {\bf 108} 137 (1993)
\bibitem{set}R Sethi, Struct. Ch. Econ. Dyn. {\bf 7} 99 (1996)
\bibitem{lm}T Lux and M Marchesi, Nature {\bf 397} 498 (1999)
\bibitem{gb}I Giardina and JP Bouchaud, Eur. Phys. J. B {\bf 31} 421
(2003)
\bibitem{Chiarella}C Chiarella, Ann. Oper. Res. {\bf 37} 101 (1992)
\bibitem{SantaFe}WB Arthur, JH Holland, B LeBaron, R Palmer and P
Tyler (1997) In: WB Arthur, SN Durlauf and DA Lane (Eds), {\it The
economy as an evolving complex system II} (Addison Wesley, Reading,
MA)
\bibitem{Flo}JD Farmer and AW Lo, Proc. Nat. Acad. Sci. {\bf 96} 9991
(1999)
\bibitem{matteo}M Marsili, Physica A {\bf 299} 93 (2001)
\bibitem{as}JV Andersen and D Sornette, Eur. Phys. J. B {\bf 31} 141 (2003)
\bibitem{f}FF Ferreira and M Marsili, Physica A {\bf 345} 657 (2005)
\bibitem{km}P Kozlowski and M Marsili, J. Phys. A {\bf 36} 11725 (2003)
\bibitem{prerc}A De Martino, I Giardina, A Tedeschi and M Marsili,
Phys. Rev. E {\bf 70} 025104(R) (2204)
 \bibitem{mmm}A De Martino, I Giardina and G Mosetti, J. Phys. A {\bf
 36} 8935 (2003)
\bibitem{cava}A Cavagna, Phys. Rev. E {\bf 59} R3783 (1999)
\end{thebibliography}
\end{document}